\RequirePackage{lineno}
\documentclass[twocolumn,aps,pra,showpacs,floatfix,superscriptaddress]{revtex4-1}
\usepackage{graphicx}
\usepackage{amsmath}
\usepackage{bm}
\usepackage{color}
\usepackage{dcolumn}
\usepackage{hyperref}

\begin{document}


\newcommand{\cmt}[1]{[\![#1]\!]}
\newcommand{\nh}{NH$^+$}
\newcommand{\gx}{$X~{^2\Pi}$}
\newcommand{\ex}{$a~{^4\Sigma^-}$}
\newcommand{\g}{${^2\Pi}$}
\newcommand{\e}{${^4\Sigma^-}$}
\newcommand{\hubers}{H\"{u}bers~\emph{et al.}}

\newcommand{\etal}{{\it et al.}}
\newcommand{\fref}[1]{Fig.~\ref{#1}}
\newcommand{\Fref}[1]{Figure \ref{#1}}
\newcommand{\sref}[1]{Sec. \ref{#1}}
\newcommand{\Eref}[1]{Eq.~(\ref{#1})}
\newcommand{\tref}[1]{Table~\ref{#1}}
\newcommand{\rtw}{\longrightarrow}
\def\veps{\varepsilon}
\newcommand{\cm}{cm$^{-1}$}

\newcommand{\NZIAS}{
Centre for Theoretical Chemistry and Physics,
New Zealand Institute for Advanced Study,
Massey University, Auckland 0745, New Zealand}
\newcommand{\UNSW}{
School of Physics, University of New South Wales, Sydney 2052, Australia}
\newcommand{\PNPI}{
Petersburg Nuclear Physics Institute, Gatchina 188300, Russia}

\title{Rotational spectrum of molecular ion NH$^+$ as a probe for $\alpha$- and $m_\mathrm{e}/m_\mathrm{p}$-variation}

\author{K. Beloy}
\affiliation{\NZIAS}

\author{M. G. Kozlov}
\affiliation{\NZIAS}
\affiliation{\PNPI}

\author{A. Borschevsky}
\affiliation{\NZIAS}

\author{A. W. Hauser}
\affiliation{\NZIAS}

\author{V. V. Flambaum}
\affiliation{\NZIAS}
\affiliation{\UNSW}

\author{P. Schwerdtfeger}
\affiliation{\NZIAS}

\date{\today}

\begin{abstract}
We identify the molecular ion NH$^+$ as a potential candidate for probing variations in the fine structure constant $\alpha$ and electron-to-proton mass ratio $\mu$. NH$^+$ has an anomalously low-lying excited {\e} state, being only a few hundred {\cm} above the ground {\g} state. Being a light molecule, this proximity is such that rotational levels of the respective states are highly intermixed for low angular momenta. We find that several low-frequency transitions within the collective 
rotational spectrum experience enhanced sensitivity to $\alpha$- and $\mu$-variation. This is attributable to the close proximity of the {\g} and {\e} states, as well as the ensuing strong spin-orbit coupling between them.
Suggestions that NH$^+$ may exist in interstellar space and recent predictions that trapped-ion precision spectroscopy will be adaptable to molecular ions make NH$^+$ a promising system for future astrophysical and laboratory studies of $\alpha$- and $\mu$-variation.
\end{abstract}


\maketitle

\section{Introduction}

The Standard Model of particle physics 
provides a solid foundation from which physical phenomena of strong and electroweak nature---from high-energy scattering to atomic and molecular structure---can be successfully described.
The Standard Model itself does not predict precise values of fundamental constants such as the fine structure constant $\alpha=e^2/\hbar c$ or the electron-to-proton mass ratio $\mu=m_\mathrm{e}/m_\mathrm{p}$, but rather accepts the experimentally observed values as input parameters to the physical theory.
Speculative theories which go beyond the Standard Model, 
such as string theories, suggest that these constants may vary in time or space \cite{Uza03}, enticing both theorists and experimentalists alike to contrive favorable means for detecting variations of these constants. 

One method for determining variations---or limitations on variations---in $\alpha$ and $\mu$ is from analysis of atomic or molecular absorption lines originating from interstellar space. 
Comparison with laboratory spectra can, in principle, reveal variations on cosmological time or distance scales. Employing this method, groups have reported 
evidence for non-zero variations in both $\alpha$ \cite{WebMurFla01etal,WebKinMur10etal} and $\mu$ \cite{ReiBunHol06etal}, though other analyses have shown no variation at similar levels of accuracy \cite{FlaKoz07b,SriChaPet04,HenMenMur09etal,Kan11,MulBeeGue10etal} (see also Comment \cite{MurWebFla07}).
Complementary to the astrophysical studies, terrestrial experiments probe $\alpha$- and $\mu$-variation on much smaller time and distance scales, but benefit from the high-precision and reproducibility 
that the spectroscopic experiments offer. Given in terms of temporal drift, laboratory experiments have placed stringent, model-free constraints on variations of both $\alpha$~\cite{RosHumSch08etal} and $\mu$~\cite{SheButCha08}: 
\begin{eqnarray}
\dot{\alpha}/\alpha&=&(-1.6\pm2.3)\times10^{-17}~\mathrm{yr}^{-1},
\nonumber\\
\dot{\mu}/\mu&=&(3.8\pm5.6)\times10^{-14}~\mathrm{yr}^{-1},
\label{Eq:lablimits}
\end{eqnarray}
with the dot signifying a derivation with respect to time.

Both the astrophysical and laboratory methods may realize significant gains by utilizing atomic or molecular species which have enhanced sensitivity to $\alpha$- and $\mu$-variation. Much theoretical effort has been dedicated to identifying such systems, with a general strategy being to locate accidental near-degeneracies within energy spectra. In this spirit, measurements have been proposed for multiply-charged ions \cite{FlaPor09,BerDzuFla10}, diatomic \cite{Fla06b,FlaKoz07a,DeMSaiSag08etal,BetUba09,Koz09,CheKan03,Dar03,ZelKotYe08} and more complex \cite{KozLapLev10,JanXuKle11,KozLev11,KozPorRei11} molecules, and even nuclei \cite{Fla06a,FlaWir09,FlaAueDmi09,LitFelDob09,HeRen07,BerDzuFla09,RelDeMGre10etal}. Note that ``near-degeneracy'' is a relative term here; for example, the ground and anomalously low-lying excited state of the $^{229}$Th nucleus---being separated by an interval of 7.6 eV \cite{BecBecBei07etal}---are nearly-degenerate relative to the typical energy scale of nuclear excitation ($\gtrsim10$ keV). 

The relative sensitivity to $\alpha$- and $\mu$-variation for a given transition may be parameterized in terms of dimensionless coefficients $Q_\alpha$ and $Q_\mu$, defined by the relation
\begin{eqnarray*}
\frac{\delta\omega}{\omega}=Q_\alpha\frac{\delta\alpha}{\alpha}+Q_\beta\frac{\delta\mu}{\mu},
\end{eqnarray*}
with $\omega$ being the transition energy. ``Typical'' transitions have sensitivity coefficients on order of unity or less. For example, in a diatomic molecule a typical fine structure transition has $Q_\alpha\approx2$, $Q_\mu\approx0$, while a typical vibrational transition has $Q_\alpha\approx0$, $Q_\mu\approx1/2$. Near-degeneracies resulting from a cancellation between fine structure and vibrational intervals can, however, lead to transitions with sensitivity coefficients orders of magnitude larger than unity~\cite{FlaKoz07a,BelBorSch10}. Several atomic, molecular, or nuclear transitions which enjoy large enhancement, however, may be irrelevant for astrophysical studies or may prove unfavorable for spectroscopic experiments, and therefore may be of limited utility.

In this paper we consider the molecular ion NH$^+$ as a candidate for measuring variation in $\alpha$ and $\mu$. NH$^+$ possesses an accidental near-degeneracy between its ground ({\gx}) and first excited ({\ex}) electronic states, these being separated by only a few hundred cm$^{-1}$ 
(for scale, the next electronic state is $\sim\!22200~\mathrm{cm}^{-1}$ higher~\cite{HubHer79}). As we will show, this near-degeneracy results in enhanced sensitivity coefficients of order 10--100 for a number of transitions within the rotational spectrum. Moreover, NH$^+$ is a light molecule which has been suggested to be a component in interstellar clouds~\cite{Fea51,ColDou68}, though to-date it has not been detected in such media~\cite{BenBruDis10etal}. Further still, motivated by the search for an electron electric dipole moment, Leanhardt~\emph{et al.}~\cite{LeaBohLoh10etal} have argued that high precision spectroscopy may be performed on molecular ions within a Paul trap, with experiments now underway for HfF$^+$~\cite{LohGraYah11DAMOPetal}. 
These considerations suggest that NH$^+$ could serve as a valuable probe of $\alpha$- and $\mu$-variation in future astrophysical or laboratory studies.

\section{Rotational Spectrum of {\nh}}
\label{Sec:rotspec}
The rotational spectrum of {\nh} was analyzed experimentally in some detail several years ago by Kawaguchi and Amano~\cite{KawAma88} and more recently by \hubers~\cite{HubEveHil09}. Both papers tabulate molecular parameters which, together with the appropriate effective Hamiltonian, are capable of furnishing the rotational spectrum. 
Evolution of the spectrum with respect to $\alpha$- and $\mu$-variation may be determined so long as the the scaling of the molecular parameters with respect to $\alpha$ and $\mu$ is known.
A list of molecular parameters relevant to the {\g} and {\e} states of {\nh} are provided in Table~\ref{Tab:params} along with their lowest-order scalings with $\alpha$ and $\mu$. 


\newcolumntype{I}{!{\vrule width 0.1pt}}
\begin{table}[t]
\caption{Molecular parameters used to describe the rotational spectrum of the {\g} and {\e} electronic states of NH$^+$ along with their lowest-order scalings with $\alpha$ and $\mu$. 
Parameters generally depend on vibrational state $v$, with some being appropriate for both electronic states (e.g., rotational constants $B$) whereas others are only appropriate for $^2\Pi$ (e.g., spin-orbit constant $A$) or $^4\Sigma^-$ (e.g., spin-spin constant $\lambda$). The physical significance of these parameters is described in Ref.~\cite{BroCar03}; the non-standard parameters $\xi_{1/2}$, $\xi_{3/2}$, and $\xi_{D}$~quantify spin-orbit coupling between $^2\Pi$ and $^4\Sigma^-$ states of similar $v$~\cite{HubEveHil09,KawAma88}.
}
\label{Tab:params}
\begin{center}
\begin{ruledtabular}
\begin{tabular}[c]{lccIclccIclc}%
Param. & Scaling 
&\multicolumn{1}{c}{\qquad}
&\multicolumn{1}{c}{\quad}
& Param. & Scaling 
&\multicolumn{1}{c}{\qquad}
&\multicolumn{1}{c}{\quad}
& Param. & Scaling \\
\hline
\\[-3mm]
$A$                 & $\alpha^2$        &&&$\lambda$       & $\alpha^2$        &&&   $T_\mathrm{nr}$     & --                \\
$B$                 & $\mu$             &&&$\lambda_D$     & $\alpha^2\mu$     &&&   $T_\mathrm{rel}$    & $\alpha^2$        \\
$D$                 & $\mu^2$           &&&$p$             & $\alpha^2\mu$     &&&   $T_\mathrm{vib}$    & $\mu^{1/2}$       \\
$H$                 & $\mu^3$           &&&$p_D$           & $\alpha^2\mu^2$   &&&   $\xi_{1/2}$         & $\alpha^2$        \\
$\gamma$            & $\alpha^2$        &&&$q$             & $\mu^2$           &&&   $\xi_{3/2}$         & $\alpha^2$        \\                       
$\gamma_D$          & $\alpha^2\mu$     &&&$q_D$           & $\mu^3$           &&&   $\xi_D$             & $\alpha^2\mu$     \\
\end{tabular}
\end{ruledtabular}
\end{center}
\end{table}

\begin{figure*}[t]
\begin{center}
\includegraphics*[scale=1]{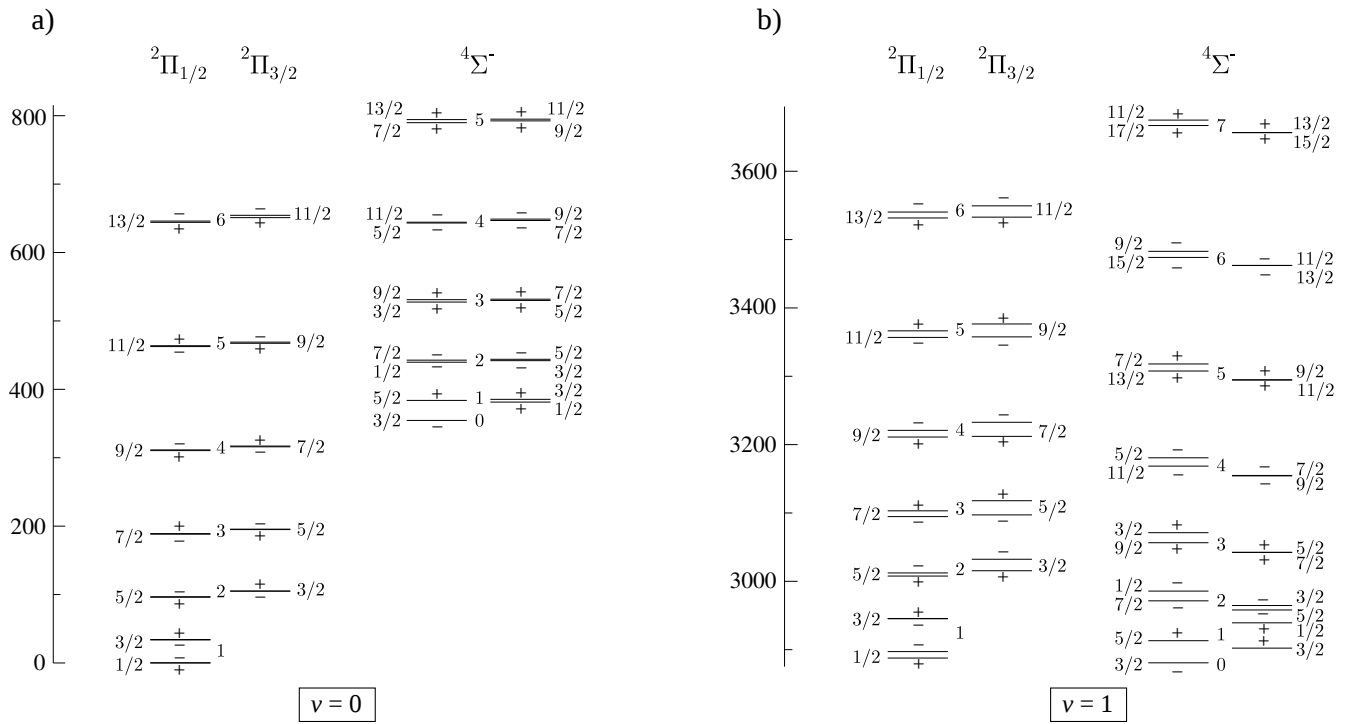}
\end{center}
\caption{Rotational spectrum of {\nh} for (a) $v=0$ and (b) $v=1$ subspaces. Integer and half-integer labels correspond to rotational and total angular momentum quantum numbers $N$ and $J$, respectively, while parity is labeled with $\pm$. Energy is in cm$^{-1}$.}
\label{Fig:spec}
\end{figure*}

Figure~\ref{Fig:spec} illustrates the rotational spectra of {\nh} for $v=0$ and $v=1$ vibrational states. 
We see that the $^4\Sigma^-,v=0$ state is in such close proximity to the ground $^2\Pi,v=0$ state that the corresponding rotational spectra begin to overlap for relatively low ($N=5$) levels of the $^2\Pi$ ladder. 
For $v=1$, the states are closer and even the lowest rotational levels are seen to be heavily intermixed.
This close proximity of {\g} and {\e} states allows for sizable coupling between the two via spin-orbit interaction. This is accounted for in the effective Hamiltonian through the non-standard molecular parameters $\xi_{1/2}$, $\xi_{3/2}$, and $\xi_{D}$ appearing in Table~\ref{Tab:params}. In general, this coupling results in noticeable perturbations to the rotational spectrum, with one example being that the ground $\Omega$-doublet interval is doubled in size~\cite{HubEveHil09}. The effect is more pronounced for close {\g} and {\e} rotational levels of similar angular momentum $J$ and parity $p$ (these being conserved quantum numbers). One noteworthy case is the $^2\Pi_{3/2},v=0,J^p=\frac{11}{2}^-$ and $^4\Sigma^-,v=0,N=4,J^p=\frac{11}{2}^-$ rotational levels (see Fig.~\ref{Fig:spec}). Due to spin-orbit coupling, the energy eigenstates are nearly equal admixtures of the unperturbed {\g} and {\e} states. 

In Table~\ref{Tab:params} we have decomposed the term energy for a given electronic and vibrational state into three contributions,
\begin{eqnarray*}
T=T_\mathrm{nr}+T_\mathrm{rel}+T_\mathrm{vib},
\end{eqnarray*}
where the respective terms correspond to the non-relativistic electronic, relativistic electronic, and vibrational contributions. For each vibrational subspace, only
$T\left({^4\Sigma^-}\right)-T\left({^2\Pi}\right)$
is required to produce the spectrum, this being supplied by Refs.~\cite{KawAma88,HubEveHil09}. However, to determine the overall $\alpha$- and $\mu$-dependence we require the partial contributions. 
The vibrational part may be inferred from Ref.~\cite{KawAma88}, and
for the $v=0$ subspace it is found to be
\begin{eqnarray}
T_\mathrm{vib}\left({^4\Sigma^-}\right)-T_\mathrm{vib}\left({^2\Pi}\right)=-179~\mathrm{cm}^{-1},
\label{Eq:Tvib}
\end{eqnarray}
corresponding to the difference in zero-point energies. For the $v=1$ subspace this difference is a factor of 3 larger. 
The relativistic electronic contribution cannot be extracted from experiment; we have determined it by \emph{ab initio} calculation to be
\begin{eqnarray*}
T_\mathrm{rel}\left({^4\Sigma^-}\right)-T_\mathrm{rel}\left({^2\Pi}\right)=41~\mathrm{cm}^{-1}.
\end{eqnarray*}
The details of this computation are reserved for the Appendix. Constraining $T\left({^4\Sigma^-}\right)-T\left({^2\Pi}\right)$ to the experimental value implies a non-relativistic electronic contribution of
\begin{eqnarray*}
T_\mathrm{nr}\left({^4\Sigma^-}\right)-T_\mathrm{nr}\left({^2\Pi}\right)=477~\mathrm{cm}^{-1}.
\end{eqnarray*}
This last part is insensitive to both $\alpha$- and $\mu$-variation.

Finally, in our analysis we choose to neglect the hyperfine splitting of the rotational levels. According to \hubers~\cite{HubEveHil09} the largest hyperfine constants are of order 100 MHz, or $3\times10^{-3}~\mathrm{cm}^{-1}$. As long as the smallest $\omega$ considered are of the order $0.1~\mathrm{cm}^{-1}$, the hyperfine interaction does not change these transition energies by more than a few percent. The same must be true for sensitivity coefficients $Q_\alpha$ and $Q_\mu$.

\begin{table*}[tbh]
\caption{$\Omega$-doublet transitions for the $v=0$ and $v=1$ vibrational states of {\nh}. 
The parity $p$ and the energy $E$ are given for the upper level of the doublet, with energy being referenced from the lowest level of the respective vibrational subspace. $E$ and $\omega$ are in {\cm} and $||E1||^2$ is in atomic units.}
  \label{Tab:OmegaDoublets}
\begin{ruledtabular}
\begin{tabular}{cD{.}{.}{4.3}D{.}{.}{2.3}D{.}{.}{4.2}D{.}{.}{4.2}D{.}{.}{1.3}cIc
                cD{.}{.}{4.3}D{.}{.}{2.3}D{.}{.}{4.2}D{.}{.}{4.2}D{.}{.}{1.3}}
$J^p$&\multicolumn{1}{c}{$E_\mathrm{up}$}
            &\multicolumn{1}{c}{$\omega$}
            &\multicolumn{1}{c}{$Q_\alpha$}
            &\multicolumn{1}{c}{$Q_\mu$}
            &\multicolumn{1}{c}{$||E1||^2$}
&\multicolumn{1}{c}{\qquad}
&\multicolumn{1}{c}{\quad}
&
$J^p$&\multicolumn{1}{c}{$E_\mathrm{up}$}
            &\multicolumn{1}{c}{$\omega$}
            &\multicolumn{1}{c}{$Q_\alpha$}
            &\multicolumn{1}{c}{$Q_\mu$}
            &\multicolumn{1}{c}{$||E1||^2$}
\\
\hline\\[-2mm]
\multicolumn{7}{c}{Transitions for $^2\Pi_{1/2},v=0$}  & \multicolumn{7}{c}{Transitions for $^2\Pi_{1/2},v=1$}\\
$ (1/2)^- $&   0.454 & 0.454 & 2.08 & 1.39 & 0.564     &&&  $ (1/2)^- $&  16.431 & 9.455 & -2.29 &  7.29 & 0.493 \\
$ (3/2)^+ $&  34.450 & 0.677 & 3.56 & 0.88 & 0.403     &&&  $ (3/2)^+ $&  64.878 & 0.418 &  0.21 & 20.94 & 0.442 \\
$ (5/2)^- $&  97.065 & 0.799 & 4.57 & 0.81 & 0.336     &&&  $ (5/2)^- $& 131.543 & 4.582 &  1.16 & -6.68 & 0.249 \\
$ (7/2)^+ $& 189.470 & 0.892 & 5.56 & 1.00 & 0.285     &&&  $ (7/2)^+ $& 222.450 & 8.464 &  2.87 & -5.13 & 0.199 \\
$ (9/2)^- $& 311.925 & 1.019 & 6.45 & 1.55 & 0.246     &&&  $ (9/2)^- $& 340.220 & 9.783 &  3.65 & -4.91 & 0.177 \\
$(11/2)^+ $& 464.416 & 1.276 & 6.80 & 2.56 & 0.215     &&&  $(11/2)^+ $& 485.857 & 9.626 &  4.21 & -4.92 & 0.162 \\
$(13/2)^- $& 646.814 & 1.854 & 6.22 & 3.98 & 0.190     &&&  $(13/2)^- $& 659.709 & 8.654 &  4.76 & -5.10 & 0.151 \\
$(15/2)^+ $& 858.917 & 3.187 & 4.83 & 5.62 & 0.168     &&&  $(15/2)^+ $& 861.832 & 7.243 &  5.46 & -5.51 & 0.142 \\
$(17/2)^- $&1100.473 & 6.435 & 3.03 & 7.44 & 0.144     &&&  $(17/2)^- $&1092.120 & 5.611 &  6.56 & -6.39 & 0.133 \\
$(19/2)^+ $&1371.183 &15.041 & 1.01 & 9.12 & 0.107     &&&  $(19/2)^+ $&1350.354 & 3.885 &  8.65 & -8.34 & 0.126 \\
$(21/2)^+ $&1699.955 &29.249 & 3.22 &-4.56 & 0.076     &&&  $(21/2)^- $&1636.231 & 2.135 & 14.22 &-13.94 & 0.119 \\
$(23/2)^- $&2020.150 &21.488 & 2.89 &-2.27 & 0.097     &&&  $(23/2)^+ $&1949.389 & 0.398 & 68.76 &-70.30 & 0.112 \\
$(25/2)^+ $&2373.578 &18.952 & 2.39 &-0.73 & 0.100     &&&  $(25/2)^+ $&2290.724 & 1.310 &-18.81 & 20.53 & 0.106 \\
$(27/2)^- $&2756.628 &18.488 & 2.00 & 0.16 & 0.097     &&&  $(27/2)^- $&2658.832 & 2.981 & -7.48 &  8.88 & 0.101 \\
$(29/2)^+ $&3167.736 &19.044 & 1.73 & 0.72 & 0.093     &&&  $(29/2)^+ $&3052.821 & 4.610 & -4.40 &  5.75 & 0.096 \\
$(31/2)^+ $&3605.981 &20.256 & 1.54 & 1.12 & 0.088     &&&  $(31/2)^- $&3472.169 & 6.194 & -3.00 &  4.34 & 0.091 \\
\\[-2mm]
\multicolumn{7}{c}{Transitions for $^2\Pi_{3/2},v=0$} &  \multicolumn{7}{c}{Transitions for $^2\Pi_{3/2},v=1$}\\
$ (3/2)^+ $& 105.405 & 0.244 & -1.97 &   2.91 & 1.592 &&& $ (3/2)^- $& 151.564 &16.803 &   1.86 & -0.67 & 1.170 \\
$ (5/2)^- $& 195.934 & 0.617 & -1.02 &   2.34 & 0.885 &&& $ (5/2)^+ $& 237.233 &20.745 &   2.76 & -1.79 & 0.664 \\
$ (7/2)^+ $& 317.542 & 1.082 & -0.55 &   2.13 & 0.598 &&& $ (7/2)^- $& 352.021 &20.618 &   3.41 & -2.37 & 0.457 \\
$ (9/2)^- $& 469.655 & 1.710 & -0.02 &   2.73 & 0.446 &&& $ (9/2)^+ $& 495.901 &18.888 &   3.91 & -2.72 & 0.353 \\
$(11/2)^- $& 655.253 & 3.410 & 19.71 & -53.15 & 0.190 &&& $(11/2)^- $& 668.781 &16.511 &   4.36 & -2.99 & 0.291 \\
$(13/2)^- $& 863.581 & 0.996 &-11.10 &  -0.22 & 0.286 &&& $(13/2)^+ $& 870.530 &13.906 &   4.86 & -3.27 & 0.251 \\
$(15/2)^+ $&1103.851 & 0.150 &-81.53 &-126.86 & 0.236 &&& $(15/2)^- $&1100.956 &11.276 &   5.50 & -3.64 & 0.221 \\
$(17/2)^+ $&1374.211 & 5.020 &  1.73 &  19.93 & 0.165 &&& $(17/2)^+ $&1359.806 & 8.724 &   6.44 & -4.25 & 0.198 \\
$(19/2)^+ $&1694.241 &20.595 &  3.65 &  -4.78 & 0.140 &&& $(19/2)^- $&1646.767 & 6.297 &   8.04 & -5.39 & 0.179 \\
$(21/2)^- $&2017.566 &15.979 &  2.66 &  -1.11 & 0.159 &&& $(21/2)^+ $&1961.476 & 4.008 &  11.39 & -7.90 & 0.164 \\
$(23/2)^+ $&2373.114 &15.516 &  2.01 &   0.32 & 0.151 &&& $(23/2)^- $&2303.527 & 1.858 &  22.27 &-16.23 & 0.150 \\
$(25/2)^- $&2757.519 &16.311 &  1.65 &   0.95 & 0.140 &&& $(25/2)^- $&2672.636 & 0.158 &-239.61 &185.81 & 0.139 \\
$(27/2)^+ $&3169.622 &17.721 &  1.43 &   1.29 & 0.130 &&& $(27/2)^+ $&3069.899 & 2.041 & -17.09 & 14.15 & 0.129 \\
$(29/2)^- $&3608.690 &19.585 &  1.30 &   1.53 & 0.120 &&& $(29/2)^- $&3492.954 & 3.787 &  -8.62 &  7.58 & 0.120 \\
\end{tabular}
\end{ruledtabular}
\end{table*}

\begin{table}[tbh]
\caption{Low-frequency ($|\omega| < 30$ \cm) transitions ${^2\Pi}\rightarrow {^4\Sigma^-}$. Negative frequencies mean that the {\e} level is below the {\g} level. {\e} levels are labeled with quantum numbers $N_J^p$ and {\g} levels are labeled with $\Omega_J^p$. $E$ and $\omega$ are in {\cm} and $||E1||^2$ is in atomic units. Transitions with $||E1||^2<10^{-3}$ a.u.~are skipped.}
 \label{tab_SigPiv0}
\begin{ruledtabular}
\begin{tabular}{ccrdddr}
$N_J^p$& $\Omega_J^p$
 &\multicolumn{1}{c}{$E_\Sigma$}
 &\multicolumn{1}{c}{$\omega$}
 &\multicolumn{1}{c}{$Q_\alpha$}
 &\multicolumn{1}{c}{$Q_\mu$}
 &\multicolumn{1}{c}{$||E1||^2$}
\\
\hline\\[-2mm]
\multicolumn{7}{c}{Transitions for $v=0$}\\
$2_{7/2}^- $& $(3/2)_{9/2}^+ $& 443.287 &-24.658&  -3.74 &  17.82 & 0.005 \\
$4_{11/2}^-$& $(1/2)_{13/2}^+$& 643.977 & -0.983& -76.84 & 249.78 & 0.002 \\
$4_{11/2}^-$& $(3/2)_{11/2}^+$& 643.977 & -7.866&  -4.68 &  30.10 & 0.165 \\
$4_{9/2}^- $& $(3/2)_{11/2}^+$& 649.676 & -2.167& -48.45 & 193.99 & 0.001
 \\[2mm]
\multicolumn{7}{c}{Transitions for $v=1$}\\
$0_{3/2}^- $& $(1/2)_{1/2}^+ $&   0.000 & -6.976& -13.68 & 22.48 & 0.385 \\
$1_{3/2}^+ $& $(1/2)_{1/2}^- $&  21.486 &  5.055&  19.73 &-33.94 & 0.208 \\
$1_{1/2}^+ $& $(1/2)_{3/2}^- $&  58.529 & -5.931& -19.95 & 25.71 & 0.222 \\
$2_{5/2}^- $& $(1/2)_{3/2}^+ $&  77.352 & 12.474&   3.48 & -8.68 & 0.023 \\
$2_{3/2}^- $& $(1/2)_{3/2}^+ $&  83.854 & 18.976&   5.44 & -6.48 & 0.003 \\
$2_{1/2}^- $& $(3/2)_{3/2}^+ $& 104.966 &-29.795&  -0.10 &  5.94 & 0.099 \\
$3_{3/2}^+ $& $(3/2)_{5/2}^- $& 190.451 &-26.037&  -1.96 &  7.87 & 0.003 \\
\end{tabular}
\end{ruledtabular}
\end{table}

\begin{table}[tbh]
\caption{Low-frequency ($|\omega| < 30$ \cm) transitions $^2\Pi_{1/2}\rightarrow {^2\Pi_{3/2}}$ in the $v=1$ subspace. Negative frequencies mean that the ${^2\Pi_{3/2}}$ level is below the ${^2\Pi_{1/2}}$ level. Levels are labeled with quantum numbers $J^p$. $E$ and $\omega$ are in {\cm} and $||E1||^2$ is in atomic units. Transitions with $||E1||^2<10^{-3}$ a.u.~are skipped.}
 \label{tab_Pi3Pi1v1}
\begin{ruledtabular}
\begin{tabular}{ccrdddr}
$(J^p)_{3/2}$& $(J^p)_{1/2}$
 &\multicolumn{1}{c}{$E_{3/2}$}
 &\multicolumn{1}{c}{$\omega$}
 &\multicolumn{1}{c}{$Q_\alpha$}
 &\multicolumn{1}{c}{$Q_\mu$}
 &\multicolumn{1}{c}{$||E1||^2$}
\\
\hline
\\[-2.5mm]
$(3/2)^+ $ &$(5/2)^- $& 134.760 &  3.217 & 30.68  &-11.53 & 0.039 \\
$(5/2)^- $ &$(7/2)^+ $& 216.488 & -5.962 & -6.89  &  0.05 & 0.020 \\
$(7/2)^+ $ &$(9/2)^- $& 331.403 & -8.817 & -0.94  & -2.13 & 0.008 \\
$(9/2)^- $ &$(11/2)^+$& 477.013 & -8.844 &  0.94  & -3.03 & 0.004 \\
$(11/2)^+$ &$(13/2)^-$& 652.270 & -7.439 &  2.18  & -3.95 & 0.002 \\
\end{tabular}
\end{ruledtabular}
\end{table}

\section{Results}
\label{Sec:Results}
Before proceeding, we must briefly discuss our management of the available experimental data. The effective Hamiltonian employed by \hubers~\cite{HubEveHil09} differs subtly from the earlier work of Kawaguchi and Amano~\cite{KawAma88}, with this difference extending to the underlying definitions---and therefore scalings---of the molecular parameters found in Table~\ref{Tab:params}. For the discussion of the previous section we chose to be consistent with \hubers, who in turn follow closely the methodical development of the effective Hamiltonian given in the book of Brown and Carrington~\cite{BroCar03}. However, in our analysis we have calculated the rotational spectrum and sensitivity coefficients by following both references as independently as possible, 
taking care to scale molecular parameters as appropriate for each case. In principle, results following from Ref.~\cite{HubEveHil09} would be preferred, as the data is more accurate and the effective Hamiltonian formulation is more transparent. Unfortunately, Ref.~\cite{HubEveHil09} lacks molecular parameters for the $v=1$ subspace and in itself does not provide information for the vibrational contribution to the term energy in the $v=0$ subspace [i.e., Eq.~(\ref{Eq:Tvib})]. 
In this section, we choose to maintain consistency by presenting our results following exclusively from Ref.~\cite{KawAma88}.
Implications of this particular choice will be discussed more in Sec.~\ref{Sec:Accuracy}.

In practice, the dimensionless sensitivity coefficients for a given transition are found from the relation (here $X=\alpha,\mu$)
\begin{eqnarray}
Q_X=\frac{\Delta q_X}{\omega},
\label{Eq:QX}
\end{eqnarray}
where $\omega=\Delta E$ is the energy difference between levels and $\Delta q_X$ is the difference between \emph{dimensional} sensitivity coefficients defined by
\begin{eqnarray}
q_X=X\frac{\partial E}{\partial X}.
\label{Eq:qX}
\end{eqnarray}
Scaling molecular parameters accordingly, we obtain rotational energy levels by diagonalizing the effective Hamiltonian for multiple values of $\mu$ and $\alpha$ in the neighborhood of the known present-day values; numerical differentiation is then used to obtain the coefficients $q_X$ for each level. Note that proper interpretation of Eq.~(\ref{Eq:qX}) requires us to specify our employed unit system, namely atomic units (i.e., the atomic unit of energy $1~\mathrm{Hartree}=\alpha^2m_\mathrm{e}c^2$ is assumed constant). A more subtle 
point is that the sensitivity coefficients $Q_X$, despite being dimensionless, also depend on this specification of atomic units. Differences $Q_X^{\prime\prime}-Q_X^{\prime}$, on the other hand, do not depend on the choice of unit system.

Tables \ref{Tab:OmegaDoublets} -- \ref{tab_Pi3Pi1v1} display our results for select (low-frequency) transitions within the {\nh} rotational spectrum. Along with the transition energy $\omega$ and sensitivity coefficients $Q_\alpha$ and $Q_\mu$, we also tabulate 
reduced matrix elements for electric dipole transition amplitudes,
\begin{eqnarray*}
  ||E1|| \equiv
   (-1)^{J'-M'}\frac{\langle J',M'|D_q|J,M\rangle}
  {\left(
  \begin{array}{ccc}
  J' & 1 & J \\
  -M'& q & M
  \end{array}\right)}\,,
\end{eqnarray*}
where $D_q$ is a spherical component of the electric dipole operator and $M$ is an angular momentum projection along the $z$-axis in a space-fixed frame. In the non-relativistic limit all off-diagonal amplitudes between electronic states $^2\Pi_{1/2}$, $^2\Pi_{3/2}$, and $^4\Sigma^-$ vanish. These amplitudes appear only after spin-orbit interaction mixes these state. As spin-orbit coupling is embedded in the effective Hamiltonian, only the dipole moments of the $^2\Pi$ and $^4\Sigma^-$ states in the molecule-fixed frame are further needed to determine the off-diagonal amplitudes. We have calculated the dipole moments for these states to be 0.91 a.u.~and 0.82 a.u., respectively (see Appendix).

\subsection{$\Omega$-doublet transitions}
 
Sensitivity of $\Omega$-doublet transitions to $\alpha$- and $\mu$-variation has been discussed previously in Refs.~\cite{Koz09,CheKan03,Dar03}. In {\nh} the situation is complicated by the proximity and strong spin-orbit coupling of the {\g} and {\e} states, resulting in less predictable behavior of the intervals. In some instances this leads to anomalously small transition energies $\omega$, with subsequently magnified sensitivity coefficients [see Eq.~(\ref{Eq:QX})].
One example is the $J=15/2$ doublet of the ${^2\Pi_{3/2}},v=0$ state. Here we find that the spin-orbit coupling leads to a more than 10-fold decrease in the transition energy, with large sensitivity coefficients $Q_\alpha=-82$ and $Q_\mu=-127$ following.
Full results for $\Omega$-doublet transitions are given in Table~\ref{Tab:OmegaDoublets}.


Another mechanism for enhancement is the state-mixing itself, as exemplified by the $J=11/2$ doublet of the ${^2\Pi_{3/2}},v=0$ state. 
Indeed, spin-orbit coupling has a significant effect on this transition energy, even causing $\omega$ to change sign. This modification of $\omega$ does not promote enhancement, however, as the absolute value of $\omega$ is found to increase by about 50\%. Enhancement here comes rather from the numerator of Eq.~(\ref{Eq:QX}). As mentioned in Sec.~\ref{Sec:rotspec}, the negative parity state in this doublet is nearly an equal admixture of {\g} and {\e} due to spin-obit coupling of near resonant levels. The {\g} portion brings no enhancement, while the {\e} portion  provides larger $\Delta q_\alpha$ and $\Delta q_\mu$ values expected of ${^2\Pi}\rightarrow{^4\Sigma^-}$ transitions.
In fact, $\Delta q_\mu$ may be readily estimated:
\begin{eqnarray*}
\Delta q_\mu&\approx&
\frac{1}{2}\left[\frac{1}{2}\left(-179~\mathrm{cm}^{-1}\right)+(-22)\left(15~\mathrm{cm}^{-1}\right)\right]
\\
&\approx&-210~\mathrm{cm}^{-1},
\end{eqnarray*}
where the first term is from a difference in vibrational energies (with a factor of 1/2 from the $\mu^{1/2}$ scaling) and the second term is from a difference in rotational energies [$B\approx15~\mathrm{cm}^{-1}$ for both states and $-22$ is the difference in $N(N+1)$]; the leading 1/2 is a weight factor accounting for the fact that the eigenstate is about ``half'' {\e}. The comparatively small 
interval, $\omega=3.4~\mathrm{cm}^{-1}$, implies a sensitivity coefficient $Q_\mu\approx-60$. This agrees with our tabulated result for this transition, wherein $Q_\alpha=20$ and $Q_\mu=-53$.
 
\subsection{{\g} -- {\e} transitions.}

From the preceding discussion, we can infer that sensitivity coefficients $Q_\alpha$ and $Q_\mu$ will be enhanced for ${^2\Pi}\rightarrow {^4\Sigma^-}$ cross transitions having $\omega$ comparable to the $\Omega$-doublet intervals ($\sim\!1-10~\mathrm{cm}^{-1}$). From the perspective of electronic transitions, these are anomalously small $\omega$. There are a number of such transitions in the spectrum, and in Tables \ref{tab_SigPiv0} we present results for transitions with $|\omega|<30~\mathrm{cm}^{-1}$. Again we find sensitivity coefficients on the order of $10-100$.

Transitions between $^2\Pi$ and $^4\Sigma^-$ states require the change of the electronic spin, $S=1/2\rightarrow S=3/2$. Thus, we would expect strong suppression of transition amplitudes. This is found to be generally true for the $v=0$ subspace. The exception is the 
transition complementary to $\Omega$-doublet transition described in the previous section (involving the other state which is a nearly equal admixture of {\g} and {\e}); this has an ``enhanced'' transition amplitude on the order of the $\Omega$-doublet transition amplitudes.
For the $v=1$ subspace, the difference in term energies is smaller ($-19~\mathrm{cm}^{-1}$ compared to $339~\mathrm{cm}^{-1}$ for $v=0$) and the mixing caused by spin-orbit interaction is generally stronger. Because of this, there are several sufficiently strong transitions. 


\subsection{Fine-structure transitions.}

Fine-structure transitions between $^2\Pi_{1/2}$ and $^2\Pi_{3/2}$ states constitute another source of low-frequency transitions; following from our analysis of $\Omega$-doubling, we may suspect enhancement of $Q_\alpha$ and $Q_\mu$ here as well. 
In the Hund's case `a' limit, these transitions require spin-flip (i.e., $\Sigma=\pm1/2 \rightarrow\Sigma=\mp1/2$, for spin-projection $\Sigma$ on the internuclear axis) and are therefore forbidden in the non-relativistic limit. The interaction with the $^4\Sigma^-$ state can open some of these transitions. It is a second order effect and the transition amplitudes are significantly smaller than for ${^2\Pi}\rightarrow {^4\Sigma^-}$ transitions. Table \ref{tab_Pi3Pi1v1} lists a number such transitions within the $v=1$ subspace, with the most sensitive transition having $Q_\alpha=31$ and $Q_\mu=-12$.

\section{Accuracy assessment}
\label{Sec:Accuracy}
As mentioned in Sec.~\ref{Sec:Results}, we analyzed the {\nh} rotational spectrum following the papers of Kawaguchi and Amano \cite{KawAma88} and \hubers~\cite{HubEveHil09} as independently as possible. A comparison of these 
results allows us to assess the accuracy of our data presented in Tables~\ref{Tab:OmegaDoublets} -- \ref{tab_Pi3Pi1v1}.
For the majority of the transitions, our values of $Q_\alpha$ and $Q_\mu$ agree to within a few percent following both references. This agreement, however, deteriorates for a handful of the transitions, with these transitions tending to be the ones 
with the largest $Q_\alpha$ and $Q_\mu$.
The bulk of the discrepancy in these cases can be attributed to a difference in 
$\omega$ [i.e., the denominator of Eq.~(\ref{Eq:QX})], which in the most extreme case differs by a factor of two. Direct measurement of the transition energies can remove this uncertainty, provided that the sensitivity coefficients are corrected appropriately,
\begin{eqnarray}
Q_X^\mathrm{corr.}=Q_X^\mathrm{theor.}\left(\frac{\omega^\mathrm{theor.}}{\omega^\mathrm{expt.}}\right).
\label{Eq:Qcorr}
\end{eqnarray}
With this prescription, theoretical uncertainty is then due to calculated $\Delta q_\alpha$ and $\Delta q_\mu$. We find that our $\Delta q_\alpha$ are consistent to $\lesssim\!10\%$, with agreement being somewhat better for $\Delta q_\mu$ at $\lesssim 5\%$.
We emphasize that our comparison here is limited to the $v=0$ subspace, as \hubers~do not give data for the $v=1$ subspace.

The above analysis effectively gauges the spread in $\Delta q_\alpha$ and $\Delta q_\mu$ arising from uncertainty in the molecular parameters of Table~\ref{Tab:params} (we note that the difference between molecular parameters tabulated in Refs.~\cite{KawAma88,HubEveHil09} is typically more than 1-$\sigma$, even after the disparity in effective Hamiltonians is accounted for~\cite{HubEveHil09}). The notable exclusion is the relativistic electronic contribution to the term energy. This is supplied by our computed value, $T_\mathrm{rel}\left({^4\Sigma^-}\right)-T_\mathrm{rel}\left({^2\Pi}\right)=41~\mathrm{cm}^{-1}$, for which we estimate an uncertainty of 10\% (see Appendix). Taking values within this window, we find deviations in $\Delta q_\alpha$ to be under 10\% for most transitions, but up to 30\% for a select few. Since $T_\mathrm{rel}$ is insensitive to variations in $\mu$, the $\Delta q_\mu$ are not affected.

Finally, we mention that our analysis is based on the lowest-order scaling of the molecular parameters. In principle, higher-order contributions alter the $\alpha$- and $\mu$-dependence of the parameters. The higher-order contributions are small, and we expect their effects on $\Delta q_\alpha$ and $\Delta q_\mu$ to be negligible. 

Altogether, we estimate uncertainty in our $\Delta q_\alpha$ and $\Delta q_\mu$ to be about 30\% and 5\%, respectively. With experimental values of $\omega$ and subsequent application of Eq.~(\ref{Eq:Qcorr}), this uncertainty extends to our tabulated sensitivity coefficients $Q_\alpha$ and $Q_\mu$. For most transitions, the 30\% estimate is highly conservative.

\section{Conclusion}

For laboratory experiments, it is useful to quantify absolute shift in addition to relative shift.
The largest $\Delta q_\alpha$ and $\Delta q_\mu$ have magnitudes of about $100~\mathrm{cm}^{-1}$ and $400~\mathrm{cm}^{-1}$, respectively. 
For temporal variations of $\alpha$ and $\mu$ at the current laboratory limits, Eq.~(\ref{Eq:lablimits}), the resulting drift in $\omega$ is
\begin{eqnarray*}
|\dot{\omega}|\approx\left(400~\mathrm{cm}^{-1}\right)\left(4\times10^{-14}~\mathrm{yr}^{-1}\right)\approx0.5~\mathrm{Hz/yr}.
\end{eqnarray*}
Here $\alpha$-variation is neglected, as it's laboratory constraint is three orders of magnitude tighter than for $\mu$-variation. We may conclude that if Hertz-level precision is achievable with molecular-ion spectroscopy, then {\nh} represents a good system to probe for $\mu$-variation in the laboratory.

The natural linewidth gives a fundamental limit to spectroscopic precision. 
The contribution to the natural linewidth from a given decay channel $n\rightarrow n^\prime$ is proportional to the product $\omega^3|\langle n|\mathbf{D}|n^\prime\rangle|^2$, with $\omega$ being the transition energy and $\langle n|\mathbf{D}|n^\prime\rangle$ being the electric dipole matrix element connecting the two states.
For the $v=0$ subspace, the dominant decay channels are the rotational transitions $N\rightarrow N-1$. As the rotational splitting $\omega_\mathrm{rot}$ grows with $N$, so too does the natural linewidth. For large $N$, where the Hund's case `b' limit is realized, we find the linewidth to be
\begin{eqnarray*}
\Gamma_\mathrm{rot}
\simeq\frac{16}{3}\left(\frac{BN}{\hbar c}\right)^{3}\mathcal{D}^2,
\end{eqnarray*}
where $\mathcal{D}$ is the dipole moment in the molecule-fixed frame. With the experimental values of $B$~\cite{HubEveHil09} and our calculated values of $\mathcal{D}$, we establish that $\Gamma_\mathrm{rot}\lesssim 30~\mathrm{Hz}$ for rotational levels $N<20$. 
For the $v=1$ subspace, decay to the ground vibrational state opens up as well. The vibrational transition energy $\omega_\mathrm{vib}$ is larger than the rotational transition energies $\omega_\mathrm{rot}$, 
whereas the dipole matrix element is suppressed in comparison. 
We find the resulting contribution to the linewidth to be
\begin{eqnarray}
\Gamma_\mathrm{vib}
\simeq\frac{4}{3}\left(\frac{\omega_\mathrm{vib}}{\hbar c}\right)^{3}\left(\frac{B}{\omega_\mathrm{vib}}\right)\chi^2\mathcal{D}^2,
\qquad
\chi\equiv \frac{d\mathrm{ln}(\mathcal{D})}{d\mathrm{ln}(R)},
\label{Eq:Gamvib}
\end{eqnarray}
$\chi$ being a factor of order unity which accounts for dependence of $\mathcal{D}$ on internuclear separation $R$ (as with $\mathcal{D}$, evaluation of $\chi$ at the equilibrium separation is implicit in the first expression).
With experimental values of $\omega_\mathrm{vib}$ \cite{KawAma88} and computed values of $\chi$ (see Appendix), we find $\Gamma_\mathrm{vib}\simeq20~\mathrm{Hz}$ for the {\g} state and $\Gamma_\mathrm{vib}\simeq85~\mathrm{Hz}$ for the {\e} state.
Assuming practical limitations $(10^{-3} - 10^{-6})\times \Gamma$, depending on
statistics, we conclude that the natural linewidth should allow measurement of the transition energies at the required level of precision.

To summarize, we have analyzed the sensitivity of the {\nh} rotational spectrum to variations in the fine structure constant $\alpha$ and electron-to-proton mass ratio $\mu$. We find enhanced sensitivity for a number of low-frequency transitions within both the $v=0$ and $v=1$ spectra, having sensitivity coefficients on the order of $10-100$. The enhanced sensitivity for these transitions is attributed to the near degeneracy of the ground {\gx} and excited {\ex} electronic states and the significant spin-orbit coupling between them. These results could prove useful in future astrophysical and laboratory searches for $\alpha$- and $\mu$-variation.

\section{Acknowledgements}
This work was supported by the Marsden Fund, administered by the Royal Society of New Zealand. MK further acknowledges support from RFBR grant 11-02-00943 and FQXi mini grant. VF further acknowledges support by the ARC. 

\section{Appendix: Computational Details}

The potential energy curves and the dipole moments of the $^{2}\Pi$ and 
$^{4}\Sigma^-$ states were obtained using the MOLPRO computational package 
\cite{MOLPRO}, within the multireference configuration interaction approach
with single and double excitations (MRCISD). Correlation consistent Dunning
aug-cc-pVQZ basis sets were employed for both atoms \cite{Dun89,KenDunHar92}%
; convergence of calculated spectroscopic constants with respect to the
basis set was verified. The second order Douglas-Kroll-Hess Hamiltonian \cite%
{DouKro74,Hes86} was used to account for scalar relativistic effects. The
contribution of spin-orbit coupling was calculated by employing the
Breit-Pauli (BP) operator. Within the MOLPRO package, the 
lowest-order one- and two-electron spin-orbit
BP operators are used for computing the matrix elements between internal configurations
(no electrons in external orbitals), while for contributions of external
configurations a mean-field one-electron Fock operator is employed. The
error introduced by this approximation is generally negligible \cite{BerSchWer00}. 

The calculated potential energy curves were used to extract the relativistic electronic contribution to the term energy, $T_\mathrm{rel}\left({^4\Sigma^-}\right)-T_\mathrm{rel}\left({^2\Pi}\right)=41~\mathrm{cm}^{-1}$. Agreement between our calculated spin-orbit splitting, $80.4~\mathrm{cm}^{-1}$, and the experimental spin-orbit constant $A=81.7~\mathrm{cm}^{-1}$ \cite{HubEveHil09} gives testament to our accuracy. Our accuracy is likely limited by higher-order ($\sim\!\alpha^4$) effects, arising primarily from the spin-orbit mixing between the $^2\Pi$ and $^4\Sigma^-$ states. This mixing is already explicitly accounted for in the effective Hamiltonian by the parameters $\xi_{1/2}$, $\xi_{3/2}$, and $\xi_{D}$, and its effect on the potential energy curves here represents somewhat of an intrusion. We estimate these effects to be a few $\mathrm{cm}^{-1}$, leading us to ascribe 10\% accuracy to our calculated value of $T_\mathrm{rel}\left({^4\Sigma^-}\right)-T_\mathrm{rel}\left({^2\Pi}\right)=41~\mathrm{cm}^{-1}$.

We have computed the dipole moments in the molecule-fixed frame (origin at the center-of-mass) to be 0.91 a.u.~and 0.82 a.u.~for the $^2\Pi$ and $^4\Sigma^-$ states, respectively. Cheng~\emph{et al.}~have previously computed the $^2\Pi$ dipole moment to be 0.7897 a.u.~\cite{CheBroRos07etal}, and we suspect that their result is more accurate.
High-accuracy is not required for the dipole moments, and this agreement is sufficient for our purposes. We also compute dependence of the dipole moments on internuclear separation; for the factor $\chi$ appearing in Eq.~(\ref{Eq:Gamvib}), we obtain values 0.73 and 2.0 for {\g} and {\e}, respectively.

%
%
%
%
%


\begin{thebibliography}{51}%
\makeatletter
\providecommand \@ifxundefined [1]{%
 \@ifx{#1\undefined}
}%
\providecommand \@ifnum [1]{%
 \ifnum #1\expandafter \@firstoftwo
 \else \expandafter \@secondoftwo
 \fi
}%
\providecommand \@ifx [1]{%
 \ifx #1\expandafter \@firstoftwo
 \else \expandafter \@secondoftwo
 \fi
}%
\providecommand \natexlab [1]{#1}%
\providecommand \enquote  [1]{``#1''}%
\providecommand \bibnamefont  [1]{#1}%
\providecommand \bibfnamefont [1]{#1}%
\providecommand \citenamefont [1]{#1}%
\providecommand \href@noop [0]{\@secondoftwo}%
\providecommand \href [0]{\begingroup \@sanitize@url \@href}%
\providecommand \@href[1]{\@@startlink{#1}\@@href}%
\providecommand \@@href[1]{\endgroup#1\@@endlink}%
\providecommand \@sanitize@url [0]{\catcode `\\12\catcode `\$12\catcode
  `\&12\catcode `\#12\catcode `\^12\catcode `\_12\catcode `\%12\relax}%
\providecommand \@@startlink[1]{}%
\providecommand \@@endlink[0]{}%
\providecommand \url  [0]{\begingroup\@sanitize@url \@url }%
\providecommand \@url [1]{\endgroup\@href {#1}{\urlprefix }}%
\providecommand \urlprefix  [0]{URL }%
\providecommand \Eprint [0]{\href }%
\providecommand \doibase [0]{http://dx.doi.org/}%
\providecommand \selectlanguage [0]{\@gobble}%
\providecommand \bibinfo  [0]{\@secondoftwo}%
\providecommand \bibfield  [0]{\@secondoftwo}%
\providecommand \translation [1]{[#1]}%
\providecommand \BibitemOpen [0]{}%
\providecommand \bibitemStop [0]{}%
\providecommand \bibitemNoStop [0]{.\EOS\space}%
\providecommand \EOS [0]{\spacefactor3000\relax}%
\providecommand \BibitemShut  [1]{\csname bibitem#1\endcsname}%
\let\auto@bib@innerbib\@empty
\bibitem [{\citenamefont {Uzan}(2003)}]{Uza03}%
  \BibitemOpen
  \bibfield  {author} {\bibinfo {author} {\bibfnamefont {J.-P.}\ \bibnamefont
  {Uzan}},\ }\href {\doibase 10.1103/RevModPhys.75.403} {\bibfield  {journal}
  {\bibinfo  {journal} {Rev. Mod. Phys.}\ }\textbf {\bibinfo {volume} {75}},\
  \bibinfo {pages} {403} (\bibinfo {year} {2003})}\BibitemShut {NoStop}%
\bibitem [{\citenamefont {{Webb~\emph{et al.}}}(2001)}]{WebMurFla01etal}%
  \BibitemOpen
  \bibfield  {author} {\bibinfo {author} {\bibfnamefont {J.~K.}\ \bibnamefont
  {{Webb~\emph{et al.}}}},\ }\href {\doibase 10.1103/PhysRevLett.87.091301}
  {\bibfield  {journal} {\bibinfo  {journal} {Phys. Rev. Lett.}\ }\textbf
  {\bibinfo {volume} {87}},\ \bibinfo {pages} {091301} (\bibinfo {year}
  {2001})}\BibitemShut {NoStop}%
\bibitem [{\citenamefont {{Webb~\emph{et al.}}}(2010)}]{WebKinMur10etal}%
  \BibitemOpen
  \bibfield  {author} {\bibinfo {author} {\bibfnamefont {J.~K.}\ \bibnamefont
  {{Webb~\emph{et al.}}}},\ }\href@noop {} {} (\bibinfo {year} {2010}),\
  \bibinfo {note} {arXiv:1008.3907v1}\BibitemShut {NoStop}%
\bibitem [{\citenamefont {{Reinhold~\emph{et al.}}}(2006)}]{ReiBunHol06etal}%
  \BibitemOpen
  \bibfield  {author} {\bibinfo {author} {\bibnamefont {{Reinhold~\emph{et
  al.}}}},\ }\href {\doibase 10.1103/PhysRevLett.96.151101} {\bibfield
  {journal} {\bibinfo  {journal} {Phys. Rev. Lett.}\ }\textbf {\bibinfo
  {volume} {96}},\ \bibinfo {pages} {151101} (\bibinfo {year}
  {2006})}\BibitemShut {NoStop}%
\bibitem [{\citenamefont {Flambaum}\ and\ \citenamefont
  {Kozlov}(2007{\natexlab{a}})}]{FlaKoz07b}%
  \BibitemOpen
  \bibfield  {author} {\bibinfo {author} {\bibfnamefont {V.~V.}\ \bibnamefont
  {Flambaum}}\ and\ \bibinfo {author} {\bibfnamefont {M.~G.}\ \bibnamefont
  {Kozlov}},\ }\href {\doibase 10.1103/PhysRevLett.98.240801} {\bibfield
  {journal} {\bibinfo  {journal} {Phys. Rev. Lett.}\ }\textbf {\bibinfo
  {volume} {98}},\ \bibinfo {pages} {240801} (\bibinfo {year}
  {2007}{\natexlab{a}})}\BibitemShut {NoStop}%
\bibitem [{\citenamefont {Srianand}\ \emph {et~al.}(2004)\citenamefont
  {Srianand}, \citenamefont {Chand}, \citenamefont {Petitjean},\ and\
  \citenamefont {Aracil}}]{SriChaPet04}%
  \BibitemOpen
  \bibfield  {author} {\bibinfo {author} {\bibfnamefont {R.}~\bibnamefont
  {Srianand}}, \bibinfo {author} {\bibfnamefont {H.}~\bibnamefont {Chand}},
  \bibinfo {author} {\bibfnamefont {P.}~\bibnamefont {Petitjean}}, \ and\
  \bibinfo {author} {\bibfnamefont {B.}~\bibnamefont {Aracil}},\ }\href
  {\doibase 10.1103/PhysRevLett.92.121302} {\bibfield  {journal} {\bibinfo
  {journal} {Phys. Rev. Lett.}\ }\textbf {\bibinfo {volume} {92}},\ \bibinfo
  {pages} {121302} (\bibinfo {year} {2004})}\BibitemShut {NoStop}%
\bibitem [{\citenamefont {{Henkel~\emph{et al.}}}(2009)}]{HenMenMur09etal}%
  \BibitemOpen
  \bibfield  {author} {\bibinfo {author} {\bibfnamefont {C.}~\bibnamefont
  {{Henkel~\emph{et al.}}}},\ }\href {\doibase 10.1051/0004-6361/200811475}
  {\bibfield  {journal} {\bibinfo  {journal} {Astron. Astrophys.}\ }\textbf
  {\bibinfo {volume} {500}},\ \bibinfo {pages} {725} (\bibinfo {year}
  {2009})}\BibitemShut {NoStop}%
\bibitem [{\citenamefont {Kanekar}(2011)}]{Kan11}%
  \BibitemOpen
  \bibfield  {author} {\bibinfo {author} {\bibfnamefont {N.}~\bibnamefont
  {Kanekar}},\ }\href@noop {} {\bibfield  {journal} {\bibinfo  {journal}
  {Astrophys. J. Lett.}\ }\textbf {\bibinfo {volume} {728}},\ \bibinfo {pages}
  {L12} (\bibinfo {year} {2011})}\BibitemShut {NoStop}%
\bibitem [{\citenamefont {{Muller~\emph{et al.}}}(2011)}]{MulBeeGue10etal}%
  \BibitemOpen
  \bibfield  {author} {\bibinfo {author} {\bibfnamefont {S.}~\bibnamefont
  {{Muller~\emph{et al.}}}},\ }\href@noop {} {} (\bibinfo {year} {2011}),\
  \bibinfo {note} {arXiv:1104.3361v1}\BibitemShut {NoStop}%
\bibitem [{\citenamefont {Murphy}\ \emph {et~al.}(2007)\citenamefont {Murphy},
  \citenamefont {Webb},\ and\ \citenamefont {Flambaum}}]{MurWebFla07}%
  \BibitemOpen
  \bibfield  {author} {\bibinfo {author} {\bibfnamefont {M.~T.}\ \bibnamefont
  {Murphy}}, \bibinfo {author} {\bibfnamefont {J.~K.}\ \bibnamefont {Webb}}, \
  and\ \bibinfo {author} {\bibfnamefont {V.~V.}\ \bibnamefont {Flambaum}},\
  }\href {\doibase 10.1103/PhysRevLett.99.239001} {\bibfield  {journal}
  {\bibinfo  {journal} {Phys. Rev. Lett.}\ }\textbf {\bibinfo {volume} {99}},\
  \bibinfo {pages} {239001} (\bibinfo {year} {2007})}\BibitemShut {NoStop}%
\bibitem [{\citenamefont {{Rosenband~\emph{et al.}}}(2008)}]{RosHumSch08etal}%
  \BibitemOpen
  \bibfield  {author} {\bibinfo {author} {\bibfnamefont {T.}~\bibnamefont
  {{Rosenband~\emph{et al.}}}},\ }\href {\doibase 10.1126/science.1154622}
  {\bibfield  {journal} {\bibinfo  {journal} {Science}\ }\textbf {\bibinfo
  {volume} {319}},\ \bibinfo {pages} {1808} (\bibinfo {year}
  {2008})}\BibitemShut {NoStop}%
\bibitem [{\citenamefont {Shelkovnikov}\ \emph {et~al.}(2008)\citenamefont
  {Shelkovnikov}, \citenamefont {Butcher}, \citenamefont {Chardonnet},\ and\
  \citenamefont {Amy-Klein}}]{SheButCha08}%
  \BibitemOpen
  \bibfield  {author} {\bibinfo {author} {\bibfnamefont {A.}~\bibnamefont
  {Shelkovnikov}}, \bibinfo {author} {\bibfnamefont {R.~J.}\ \bibnamefont
  {Butcher}}, \bibinfo {author} {\bibfnamefont {C.}~\bibnamefont {Chardonnet}},
  \ and\ \bibinfo {author} {\bibfnamefont {A.}~\bibnamefont {Amy-Klein}},\
  }\href {\doibase 10.1103/PhysRevLett.100.150801} {\bibfield  {journal}
  {\bibinfo  {journal} {Phys. Rev. Lett.}\ }\textbf {\bibinfo {volume} {100}},\
  \bibinfo {pages} {150801} (\bibinfo {year} {2008})}\BibitemShut {NoStop}%
\bibitem [{\citenamefont {Flambaum}\ and\ \citenamefont
  {Porsev}(2009)}]{FlaPor09}%
  \BibitemOpen
  \bibfield  {author} {\bibinfo {author} {\bibfnamefont {V.~V.}\ \bibnamefont
  {Flambaum}}\ and\ \bibinfo {author} {\bibfnamefont {S.~G.}\ \bibnamefont
  {Porsev}},\ }\href {\doibase 10.1103/PhysRevA.80.064502} {\bibfield
  {journal} {\bibinfo  {journal} {Phys. Rev. A}\ }\textbf {\bibinfo {volume}
  {80}},\ \bibinfo {pages} {064502} (\bibinfo {year} {2009})}\BibitemShut
  {NoStop}%
\bibitem [{\citenamefont {Berengut}\ \emph {et~al.}(2010)\citenamefont
  {Berengut}, \citenamefont {Dzuba},\ and\ \citenamefont
  {Flambaum}}]{BerDzuFla10}%
  \BibitemOpen
  \bibfield  {author} {\bibinfo {author} {\bibfnamefont {J.~C.}\ \bibnamefont
  {Berengut}}, \bibinfo {author} {\bibfnamefont {V.~A.}\ \bibnamefont {Dzuba}},
  \ and\ \bibinfo {author} {\bibfnamefont {V.~V.}\ \bibnamefont {Flambaum}},\
  }\href {\doibase 10.1103/PhysRevLett.105.120801} {\bibfield  {journal}
  {\bibinfo  {journal} {Phys. Rev. Lett.}\ }\textbf {\bibinfo {volume} {105}},\
  \bibinfo {pages} {120801} (\bibinfo {year} {2010})}\BibitemShut {NoStop}%
\bibitem [{\citenamefont {Flambaum}(2006{\natexlab{a}})}]{Fla06b}%
  \BibitemOpen
  \bibfield  {author} {\bibinfo {author} {\bibfnamefont {V.~V.}\ \bibnamefont
  {Flambaum}},\ }\href {\doibase 10.1103/PhysRevA.73.034101} {\bibfield
  {journal} {\bibinfo  {journal} {Phys. Rev. A}\ }\textbf {\bibinfo {volume}
  {73}},\ \bibinfo {pages} {034101} (\bibinfo {year}
  {2006}{\natexlab{a}})}\BibitemShut {NoStop}%
\bibitem [{\citenamefont {Flambaum}\ and\ \citenamefont
  {Kozlov}(2007{\natexlab{b}})}]{FlaKoz07a}%
  \BibitemOpen
  \bibfield  {author} {\bibinfo {author} {\bibfnamefont {V.~V.}\ \bibnamefont
  {Flambaum}}\ and\ \bibinfo {author} {\bibfnamefont {M.~G.}\ \bibnamefont
  {Kozlov}},\ }\href {\doibase 10.1103/PhysRevLett.99.150801} {\bibfield
  {journal} {\bibinfo  {journal} {Phys. Rev. Lett.}\ }\textbf {\bibinfo
  {volume} {99}},\ \bibinfo {pages} {150801} (\bibinfo {year}
  {2007}{\natexlab{b}})}\BibitemShut {NoStop}%
\bibitem [{\citenamefont {{DeMille~\emph{et al.}}}(2008)}]{DeMSaiSag08etal}%
  \BibitemOpen
  \bibfield  {author} {\bibinfo {author} {\bibfnamefont {D.}~\bibnamefont
  {{DeMille~\emph{et al.}}}},\ }\href {\doibase 10.1103/PhysRevLett.100.043202}
  {\bibfield  {journal} {\bibinfo  {journal} {Phys. Rev. Lett.}\ }\textbf
  {\bibinfo {volume} {100}},\ \bibinfo {pages} {043202} (\bibinfo {year}
  {2008})}\BibitemShut {NoStop}%
\bibitem [{\citenamefont {Bethlem}\ and\ \citenamefont
  {Ubachs}(2009)}]{BetUba09}%
  \BibitemOpen
  \bibfield  {author} {\bibinfo {author} {\bibfnamefont {H.~L.}\ \bibnamefont
  {Bethlem}}\ and\ \bibinfo {author} {\bibfnamefont {W.}~\bibnamefont
  {Ubachs}},\ }\href@noop {} {\bibfield  {journal} {\bibinfo  {journal}
  {Faraday Discuss.}\ }\textbf {\bibinfo {volume} {142}},\ \bibinfo {pages}
  {25} (\bibinfo {year} {2009})}\BibitemShut {NoStop}%
\bibitem [{\citenamefont {Kozlov}(2009)}]{Koz09}%
  \BibitemOpen
  \bibfield  {author} {\bibinfo {author} {\bibfnamefont {M.~G.}\ \bibnamefont
  {Kozlov}},\ }\href {\doibase 10.1103/PhysRevA.80.022118} {\bibfield
  {journal} {\bibinfo  {journal} {Phys. Rev. A}\ }\textbf {\bibinfo {volume}
  {80}},\ \bibinfo {pages} {022118} (\bibinfo {year} {2009})}\BibitemShut
  {NoStop}%
\bibitem [{\citenamefont {Chengalur}\ and\ \citenamefont
  {Kanekar}(2003)}]{CheKan03}%
  \BibitemOpen
  \bibfield  {author} {\bibinfo {author} {\bibfnamefont {J.~N.}\ \bibnamefont
  {Chengalur}}\ and\ \bibinfo {author} {\bibfnamefont {N.}~\bibnamefont
  {Kanekar}},\ }\href {\doibase 10.1103/PhysRevLett.91.241302} {\bibfield
  {journal} {\bibinfo  {journal} {Phys. Rev. Lett.}\ }\textbf {\bibinfo
  {volume} {91}},\ \bibinfo {pages} {241302} (\bibinfo {year}
  {2003})}\BibitemShut {NoStop}%
\bibitem [{\citenamefont {Darling}(2003)}]{Dar03}%
  \BibitemOpen
  \bibfield  {author} {\bibinfo {author} {\bibfnamefont {J.}~\bibnamefont
  {Darling}},\ }\href {\doibase 10.1103/PhysRevLett.91.011301} {\bibfield
  {journal} {\bibinfo  {journal} {Phys. Rev. Lett.}\ }\textbf {\bibinfo
  {volume} {91}},\ \bibinfo {pages} {011301} (\bibinfo {year}
  {2003})}\BibitemShut {NoStop}%
\bibitem [{\citenamefont {Zelevinsky}\ \emph {et~al.}(2008)\citenamefont
  {Zelevinsky}, \citenamefont {Kotochigova},\ and\ \citenamefont
  {Ye}}]{ZelKotYe08}%
  \BibitemOpen
  \bibfield  {author} {\bibinfo {author} {\bibfnamefont {T.}~\bibnamefont
  {Zelevinsky}}, \bibinfo {author} {\bibfnamefont {S.}~\bibnamefont
  {Kotochigova}}, \ and\ \bibinfo {author} {\bibfnamefont {J.}~\bibnamefont
  {Ye}},\ }\href {\doibase 10.1103/PhysRevLett.100.043201} {\bibfield
  {journal} {\bibinfo  {journal} {Phys. Rev. Lett.}\ }\textbf {\bibinfo
  {volume} {100}},\ \bibinfo {pages} {043201} (\bibinfo {year}
  {2008})}\BibitemShut {NoStop}%
\bibitem [{\citenamefont {Kozlov}\ \emph {et~al.}(2010)\citenamefont {Kozlov},
  \citenamefont {Lapinov},\ and\ \citenamefont {Levshakov}}]{KozLapLev10}%
  \BibitemOpen
  \bibfield  {author} {\bibinfo {author} {\bibfnamefont {M.~G.}\ \bibnamefont
  {Kozlov}}, \bibinfo {author} {\bibfnamefont {A.~V.}\ \bibnamefont {Lapinov}},
  \ and\ \bibinfo {author} {\bibfnamefont {S.~A.}\ \bibnamefont {Levshakov}},\
  }\href@noop {} {\bibfield  {journal} {\bibinfo  {journal} {J. Phys. B}\
  }\textbf {\bibinfo {volume} {43}},\ \bibinfo {pages} {074003} (\bibinfo
  {year} {2010})}\BibitemShut {NoStop}%
\bibitem [{\citenamefont {Jansen}\ \emph {et~al.}(2011)\citenamefont {Jansen},
  \citenamefont {Xu}, \citenamefont {Kleiner}, \citenamefont {Ubachs},\ and\
  \citenamefont {Bethlem}}]{JanXuKle11}%
  \BibitemOpen
  \bibfield  {author} {\bibinfo {author} {\bibfnamefont {P.}~\bibnamefont
  {Jansen}}, \bibinfo {author} {\bibfnamefont {L.}~\bibnamefont {Xu}}, \bibinfo
  {author} {\bibfnamefont {I.}~\bibnamefont {Kleiner}}, \bibinfo {author}
  {\bibfnamefont {W.}~\bibnamefont {Ubachs}}, \ and\ \bibinfo {author}
  {\bibfnamefont {H.~L.}\ \bibnamefont {Bethlem}},\ }\href {\doibase
  10.1103/PhysRevLett.106.100801} {\bibfield  {journal} {\bibinfo  {journal}
  {Phys. Rev. Lett.}\ }\textbf {\bibinfo {volume} {106}},\ \bibinfo {pages}
  {100801} (\bibinfo {year} {2011})}\BibitemShut {NoStop}%
\bibitem [{\citenamefont {Kozlov}\ and\ \citenamefont
  {Levshakov}(2011)}]{KozLev11}%
  \BibitemOpen
  \bibfield  {author} {\bibinfo {author} {\bibfnamefont {M.~G.}\ \bibnamefont
  {Kozlov}}\ and\ \bibinfo {author} {\bibfnamefont {S.~A.}\ \bibnamefont
  {Levshakov}},\ }\href@noop {} {\bibfield  {journal} {\bibinfo  {journal}
  {Astrophys. J.}\ }\textbf {\bibinfo {volume} {726}},\ \bibinfo {pages} {65}
  (\bibinfo {year} {2011})}\BibitemShut {NoStop}%
\bibitem [{\citenamefont {Kozlov}\ \emph {et~al.}(2011)\citenamefont {Kozlov},
  \citenamefont {Porsev},\ and\ \citenamefont {Reimers}}]{KozPorRei11}%
  \BibitemOpen
  \bibfield  {author} {\bibinfo {author} {\bibfnamefont {M.~G.}\ \bibnamefont
  {Kozlov}}, \bibinfo {author} {\bibfnamefont {S.~G.}\ \bibnamefont {Porsev}},
  \ and\ \bibinfo {author} {\bibfnamefont {D.}~\bibnamefont {Reimers}},\
  }\href@noop {} {} (\bibinfo {year} {2011}),\ \bibinfo {note}
  {arXiv:1103.4739}\BibitemShut {NoStop}%
\bibitem [{\citenamefont {Flambaum}(2006{\natexlab{b}})}]{Fla06a}%
  \BibitemOpen
  \bibfield  {author} {\bibinfo {author} {\bibfnamefont {V.~V.}\ \bibnamefont
  {Flambaum}},\ }\href {\doibase 10.1103/PhysRevLett.97.092502} {\bibfield
  {journal} {\bibinfo  {journal} {Phys. Rev. Lett.}\ }\textbf {\bibinfo
  {volume} {97}},\ \bibinfo {pages} {092502} (\bibinfo {year}
  {2006}{\natexlab{b}})}\BibitemShut {NoStop}%
\bibitem [{\citenamefont {Flambaum}\ and\ \citenamefont
  {Wiringa}(2009)}]{FlaWir09}%
  \BibitemOpen
  \bibfield  {author} {\bibinfo {author} {\bibfnamefont {V.~V.}\ \bibnamefont
  {Flambaum}}\ and\ \bibinfo {author} {\bibfnamefont {R.~B.}\ \bibnamefont
  {Wiringa}},\ }\href {\doibase 10.1103/PhysRevC.79.034302} {\bibfield
  {journal} {\bibinfo  {journal} {Phys. Rev. C}\ }\textbf {\bibinfo {volume}
  {79}},\ \bibinfo {pages} {034302} (\bibinfo {year} {2009})}\BibitemShut
  {NoStop}%
\bibitem [{\citenamefont {Flambaum}\ \emph {et~al.}(2009)\citenamefont
  {Flambaum}, \citenamefont {Auerbach},\ and\ \citenamefont
  {Dmitriev}}]{FlaAueDmi09}%
  \BibitemOpen
  \bibfield  {author} {\bibinfo {author} {\bibfnamefont {V.~V.}\ \bibnamefont
  {Flambaum}}, \bibinfo {author} {\bibfnamefont {N.}~\bibnamefont {Auerbach}},
  \ and\ \bibinfo {author} {\bibfnamefont {V.~F.}\ \bibnamefont {Dmitriev}},\
  }\href@noop {} {\bibfield  {journal} {\bibinfo  {journal} {EPL}\ }\textbf
  {\bibinfo {volume} {85}},\ \bibinfo {pages} {50005} (\bibinfo {year}
  {2009})}\BibitemShut {NoStop}%
\bibitem [{\citenamefont {Litvinova}\ \emph {et~al.}(2009)\citenamefont
  {Litvinova}, \citenamefont {Feldmeier}, \citenamefont {Dobaczewski},\ and\
  \citenamefont {Flambaum}}]{LitFelDob09}%
  \BibitemOpen
  \bibfield  {author} {\bibinfo {author} {\bibfnamefont {E.}~\bibnamefont
  {Litvinova}}, \bibinfo {author} {\bibfnamefont {H.}~\bibnamefont
  {Feldmeier}}, \bibinfo {author} {\bibfnamefont {J.}~\bibnamefont
  {Dobaczewski}}, \ and\ \bibinfo {author} {\bibfnamefont {V.}~\bibnamefont
  {Flambaum}},\ }\href {\doibase 10.1103/PhysRevC.79.064303} {\bibfield
  {journal} {\bibinfo  {journal} {Phys. Rev. C}\ }\textbf {\bibinfo {volume}
  {79}},\ \bibinfo {pages} {064303} (\bibinfo {year} {2009})}\BibitemShut
  {NoStop}%
\bibitem [{\citenamefont {He}\ and\ \citenamefont {Ren}(2007)}]{HeRen07}%
  \BibitemOpen
  \bibfield  {author} {\bibinfo {author} {\bibfnamefont {X.}~\bibnamefont
  {He}}\ and\ \bibinfo {author} {\bibfnamefont {Z.}~\bibnamefont {Ren}},\
  }\href@noop {} {\bibfield  {journal} {\bibinfo  {journal} {J. Phys. G}\
  }\textbf {\bibinfo {volume} {34}},\ \bibinfo {pages} {1611} (\bibinfo {year}
  {2007})}\BibitemShut {NoStop}%
\bibitem [{\citenamefont {Berengut}\ \emph {et~al.}(2009)\citenamefont
  {Berengut}, \citenamefont {Dzuba}, \citenamefont {Flambaum},\ and\
  \citenamefont {Porsev}}]{BerDzuFla09}%
  \BibitemOpen
  \bibfield  {author} {\bibinfo {author} {\bibfnamefont {J.~C.}\ \bibnamefont
  {Berengut}}, \bibinfo {author} {\bibfnamefont {V.~A.}\ \bibnamefont {Dzuba}},
  \bibinfo {author} {\bibfnamefont {V.~V.}\ \bibnamefont {Flambaum}}, \ and\
  \bibinfo {author} {\bibfnamefont {S.~G.}\ \bibnamefont {Porsev}},\ }\href
  {\doibase 10.1103/PhysRevLett.102.210801} {\bibfield  {journal} {\bibinfo
  {journal} {Phys. Rev. Lett.}\ }\textbf {\bibinfo {volume} {102}},\ \bibinfo
  {pages} {210801} (\bibinfo {year} {2009})}\BibitemShut {NoStop}%
\bibitem [{\citenamefont {{Rellergert~\emph{et al.}}}(2010)}]{RelDeMGre10etal}%
  \BibitemOpen
  \bibfield  {author} {\bibinfo {author} {\bibfnamefont {W.~G.}\ \bibnamefont
  {{Rellergert~\emph{et al.}}}},\ }\href {\doibase
  10.1103/PhysRevLett.104.200802} {\bibfield  {journal} {\bibinfo  {journal}
  {Phys. Rev. Lett.}\ }\textbf {\bibinfo {volume} {104}},\ \bibinfo {pages}
  {200802} (\bibinfo {year} {2010})}\BibitemShut {NoStop}%
\bibitem [{\citenamefont {{Beck~\emph{et al.}}}(2007)}]{BecBecBei07etal}%
  \BibitemOpen
  \bibfield  {author} {\bibinfo {author} {\bibfnamefont {B.~R.}\ \bibnamefont
  {{Beck~\emph{et al.}}}},\ }\href {\doibase 10.1103/PhysRevLett.98.142501}
  {\bibfield  {journal} {\bibinfo  {journal} {Phys. Rev. Lett.}\ }\textbf
  {\bibinfo {volume} {98}},\ \bibinfo {pages} {142501} (\bibinfo {year}
  {2007})}\BibitemShut {NoStop}%
\bibitem [{\citenamefont {Beloy}\ \emph {et~al.}(2010)\citenamefont {Beloy},
  \citenamefont {Borschevsky}, \citenamefont {Schwerdtfeger},\ and\
  \citenamefont {Flambaum}}]{BelBorSch10}%
  \BibitemOpen
  \bibfield  {author} {\bibinfo {author} {\bibfnamefont {K.}~\bibnamefont
  {Beloy}}, \bibinfo {author} {\bibfnamefont {A.}~\bibnamefont {Borschevsky}},
  \bibinfo {author} {\bibfnamefont {P.}~\bibnamefont {Schwerdtfeger}}, \ and\
  \bibinfo {author} {\bibfnamefont {V.~V.}\ \bibnamefont {Flambaum}},\ }\href
  {\doibase 10.1103/PhysRevA.82.022106} {\bibfield  {journal} {\bibinfo
  {journal} {Phys. Rev. A}\ }\textbf {\bibinfo {volume} {82}},\ \bibinfo
  {pages} {022106} (\bibinfo {year} {2010})}\BibitemShut {NoStop}%
\bibitem [{\citenamefont {Huber}\ and\ \citenamefont
  {Herzberg}(1979)}]{HubHer79}%
  \BibitemOpen
  \bibfield  {author} {\bibinfo {author} {\bibfnamefont {K.~P.}\ \bibnamefont
  {Huber}}\ and\ \bibinfo {author} {\bibfnamefont {G.}~\bibnamefont
  {Herzberg}},\ }\href@noop {} {\emph {\bibinfo {title} {Molecular Spectra and
  Molecular Structure, Vol. IV: Constants of Diatomic Molecules}}}\ (\bibinfo
  {publisher} {Van Nostrand Reinhold Company Inc.},\ \bibinfo {address} {New
  York, NY, USA},\ \bibinfo {year} {1979})\BibitemShut {NoStop}%
\bibitem [{\citenamefont {Feast}(1951)}]{Fea51}%
  \BibitemOpen
  \bibfield  {author} {\bibinfo {author} {\bibfnamefont {M.~W.}\ \bibnamefont
  {Feast}},\ }\href@noop {} {\bibfield  {journal} {\bibinfo  {journal}
  {Astrophys. J.}\ }\textbf {\bibinfo {volume} {114}},\ \bibinfo {pages} {344}
  (\bibinfo {year} {1951})}\BibitemShut {NoStop}%
\bibitem [{\citenamefont {Colin}\ and\ \citenamefont
  {Douglas}(1968)}]{ColDou68}%
  \BibitemOpen
  \bibfield  {author} {\bibinfo {author} {\bibfnamefont {R.}~\bibnamefont
  {Colin}}\ and\ \bibinfo {author} {\bibfnamefont {A.~E.}\ \bibnamefont
  {Douglas}},\ }\href@noop {} {\bibfield  {journal} {\bibinfo  {journal} {Can.
  J. Phys.}\ }\textbf {\bibinfo {volume} {46}},\ \bibinfo {pages} {61}
  (\bibinfo {year} {1968})}\BibitemShut {NoStop}%
\bibitem [{\citenamefont {{Benz~\emph{et al.}}}(2010)}]{BenBruDis10etal}%
  \BibitemOpen
  \bibfield  {author} {\bibinfo {author} {\bibfnamefont {A.~O.}\ \bibnamefont
  {{Benz~\emph{et al.}}}},\ }\href@noop {} {\bibfield  {journal} {\bibinfo
  {journal} {Astronom. Astrophys.}\ }\textbf {\bibinfo {volume} {521}},\
  \bibinfo {pages} {L35} (\bibinfo {year} {2010})}\BibitemShut {NoStop}%
\bibitem [{\citenamefont {{Leanhardt~\emph{et al.}}}(2010)}]{LeaBohLoh10etal}%
  \BibitemOpen
  \bibfield  {author} {\bibinfo {author} {\bibfnamefont {A.~E.}\ \bibnamefont
  {{Leanhardt~\emph{et al.}}}},\ }\href@noop {} {} (\bibinfo {year} {2010}),\
  \bibinfo {note} {arXiv:1008.2997v2}\BibitemShut {NoStop}%
\bibitem [{\citenamefont {{Loh~\emph{et al.}}}(2011)}]{LohGraYah11DAMOPetal}%
  \BibitemOpen
  \bibfield  {author} {\bibinfo {author} {\bibfnamefont {H.}~\bibnamefont
  {{Loh~\emph{et al.}}}},\ }\href@noop {} {} (\bibinfo {year} {2011}),\
  \bibinfo {note} {{Bulletin of the American Physical Society, \emph{42nd
  Annual Meeting of the APS Division of Atomic, Molecular and Optical
  Physics}}, Atlanta, GA, USA}\BibitemShut {NoStop}%
\bibitem [{\citenamefont {Kawaguchi}\ and\ \citenamefont
  {Amano}(1988)}]{KawAma88}%
  \BibitemOpen
  \bibfield  {author} {\bibinfo {author} {\bibfnamefont {K.}~\bibnamefont
  {Kawaguchi}}\ and\ \bibinfo {author} {\bibfnamefont {T.}~\bibnamefont
  {Amano}},\ }\href@noop {} {\bibfield  {journal} {\bibinfo  {journal} {J.
  Chem. Phys.}\ }\textbf {\bibinfo {volume} {88}},\ \bibinfo {pages} {15}
  (\bibinfo {year} {1988})}\BibitemShut {NoStop}%
\bibitem [{\citenamefont {H\"{u}bers}\ \emph {et~al.}(2009)\citenamefont
  {H\"{u}bers}, \citenamefont {Evenson}, \citenamefont {Hill},\ and\
  \citenamefont {Brown}}]{HubEveHil09}%
  \BibitemOpen
  \bibfield  {author} {\bibinfo {author} {\bibfnamefont {H.}~\bibnamefont
  {H\"{u}bers}}, \bibinfo {author} {\bibfnamefont {K.~M.}\ \bibnamefont
  {Evenson}}, \bibinfo {author} {\bibfnamefont {C.}~\bibnamefont {Hill}}, \
  and\ \bibinfo {author} {\bibfnamefont {J.~M.}\ \bibnamefont {Brown}},\
  }\href@noop {} {\bibfield  {journal} {\bibinfo  {journal} {J. Chem. Phys.}\
  }\textbf {\bibinfo {volume} {131}},\ \bibinfo {pages} {034311} (\bibinfo
  {year} {2009})}\BibitemShut {NoStop}%
\bibitem [{\citenamefont {Brown}\ and\ \citenamefont
  {Carrington}(2003)}]{BroCar03}%
  \BibitemOpen
  \bibfield  {author} {\bibinfo {author} {\bibfnamefont {J.}~\bibnamefont
  {Brown}}\ and\ \bibinfo {author} {\bibfnamefont {A.}~\bibnamefont
  {Carrington}},\ }\href@noop {} {\emph {\bibinfo {title} {Rotational
  Spectroscopy of Diatomic Molecules}}}\ (\bibinfo  {publisher} {Cambridge
  University Press},\ \bibinfo {address} {Cambridge, UK},\ \bibinfo {year}
  {2003})\BibitemShut {NoStop}%
\bibitem [{\citenamefont {Werner}\ \emph {et~al.}()\citenamefont {Werner} \emph
  {et~al.}}]{MOLPRO}%
  \BibitemOpen
  \bibfield  {author} {\bibinfo {author} {\bibfnamefont {H.-J.}\ \bibnamefont
  {Werner}} \emph {et~al.},\ }\href@noop {} {\enquote {\bibinfo {title}
  {{MOLPRO}, a package of \emph{ab initio} programs, version 2009.1},}\
  }\bibinfo {note} {{http://www.molpro.net}}\BibitemShut {NoStop}%
\bibitem [{\citenamefont {{Dunning, Jr.}}(1989)}]{Dun89}%
  \BibitemOpen
  \bibfield  {author} {\bibinfo {author} {\bibfnamefont {T.~H.}\ \bibnamefont
  {{Dunning, Jr.}}},\ }\href {\doibase 10.1103/PhysRevD.67.083507} {\bibfield
  {journal} {\bibinfo  {journal} {J. Chem. Phys.}\ }\textbf {\bibinfo {volume}
  {90}},\ \bibinfo {pages} {1007} (\bibinfo {year} {1989})}\BibitemShut
  {NoStop}%
\bibitem [{\citenamefont {Kendall}\ \emph {et~al.}(1992)\citenamefont
  {Kendall}, \citenamefont {{Dunning, Jr.}},\ and\ \citenamefont
  {Harrison}}]{KenDunHar92}%
  \BibitemOpen
  \bibfield  {author} {\bibinfo {author} {\bibfnamefont {R.~A.}\ \bibnamefont
  {Kendall}}, \bibinfo {author} {\bibfnamefont {T.~H.}\ \bibnamefont {{Dunning,
  Jr.}}}, \ and\ \bibinfo {author} {\bibfnamefont {R.~J.}\ \bibnamefont
  {Harrison}},\ }\href {\doibase 10.1063/1.478678} {\bibfield  {journal}
  {\bibinfo  {journal} {J. Chem. Phys.}\ }\textbf {\bibinfo {volume} {110}},\
  \bibinfo {pages} {6796} (\bibinfo {year} {1992})}\BibitemShut {NoStop}%
\bibitem [{\citenamefont {Douglas}\ and\ \citenamefont
  {Kroll}(1974)}]{DouKro74}%
  \BibitemOpen
  \bibfield  {author} {\bibinfo {author} {\bibfnamefont {M.}~\bibnamefont
  {Douglas}}\ and\ \bibinfo {author} {\bibfnamefont {N.~M.}\ \bibnamefont
  {Kroll}},\ }\href {\doibase DOI: 10.1016/0003-4916(74)90333-9} {\bibfield
  {journal} {\bibinfo  {journal} {Ann. Phys.}\ }\textbf {\bibinfo {volume}
  {82}},\ \bibinfo {pages} {89 } (\bibinfo {year} {1974})}\BibitemShut
  {NoStop}%
\bibitem [{\citenamefont {Hess}(1986)}]{Hes86}%
  \BibitemOpen
  \bibfield  {author} {\bibinfo {author} {\bibfnamefont {B.~A.}\ \bibnamefont
  {Hess}},\ }\href {\doibase 10.1103/PhysRevA.33.3742} {\bibfield  {journal}
  {\bibinfo  {journal} {Phys. Rev. A}\ }\textbf {\bibinfo {volume} {33}},\
  \bibinfo {pages} {3742} (\bibinfo {year} {1986})}\BibitemShut {NoStop}%
\bibitem [{\citenamefont {Berning}\ \emph {et~al.}(2000)\citenamefont
  {Berning}, \citenamefont {Schweizer}, \citenamefont {Werner}, \citenamefont
  {Knowles},\ and\ \citenamefont {Palmieri}}]{BerSchWer00}%
  \BibitemOpen
  \bibfield  {author} {\bibinfo {author} {\bibfnamefont {A.}~\bibnamefont
  {Berning}}, \bibinfo {author} {\bibfnamefont {M.}~\bibnamefont {Schweizer}},
  \bibinfo {author} {\bibfnamefont {H.}~\bibnamefont {Werner}}, \bibinfo
  {author} {\bibfnamefont {P.}~\bibnamefont {Knowles}}, \ and\ \bibinfo
  {author} {\bibfnamefont {P.}~\bibnamefont {Palmieri}},\ }\href@noop {}
  {\bibfield  {journal} {\bibinfo  {journal} {Mol. Phys.}\ }\textbf {\bibinfo
  {volume} {98}},\ \bibinfo {pages} {1823} (\bibinfo {year}
  {2000})}\BibitemShut {NoStop}%
\bibitem [{\citenamefont {Cheng}\ \emph {et~al.}(2007)\citenamefont {Cheng}
  \emph {et~al.}}]{CheBroRos07etal}%
  \BibitemOpen
  \bibfield  {author} {\bibinfo {author} {\bibfnamefont {M.}~\bibnamefont
  {Cheng}} \emph {et~al.},\ }\href {\doibase 10.1103/PhysRevA.75.012502}
  {\bibfield  {journal} {\bibinfo  {journal} {Phys. Rev. A}\ }\textbf {\bibinfo
  {volume} {75}},\ \bibinfo {pages} {012502} (\bibinfo {year}
  {2007})}\BibitemShut {NoStop}%
\end{thebibliography}

%

\end{document}